\documentclass[12pt,preprint]{aastex}

\def\phil{\hbox{$\phi^{\rm i}_{\rm l}$}}
\def\phit{\hbox{$\phi^{\rm i}_{\rm t}$}}
\def\gammai{\hbox{$\Gamma^{\rm i}$}}
\def\etai{\hbox{$\eta^{\rm i}$}}
\def\remi{\hbox{$r^{\rm i}_{\rm em}$}}
\def\sii{\hbox{$s^{\rm i}_{\rm L}$}}
\shorttitle{Emission geometry of PSR~B0329+54}
\shortauthors{Gangadhara and Gupta}
\received{}

\begin{document}
\title{Understanding the radio emission geometry of \\ PSR~B0329+54 }

\author{R. T. Gangadhara\altaffilmark{1} and Y. Gupta\altaffilmark{2}}

\affil{\altaffilmark{1}Indian Institute of Astrophysics, Bangalore-560034, India}

\affil{\altaffilmark{2}National Centre for Radio Astrophysics, TIFR,\\ Pune University Campus, 
Pune--411007, India}
\altaffiltext{1}{E-mail: {\tt ganga@iiap.ernet.in}}
\altaffiltext{2}{E-mail: {\tt ygupta@ncra.tifr.res.in }}

\begin{abstract}
We have analyzed high-quality single pulse data of PSR~B0329+54 at
325~MHz and 606~MHz to study the structure of the emission beam. Using
the window--threshold technique, which is suitable for detecting weak 
emission components, we have detected 4 additional emission components 
in the pulse window. Three of these are new components and the fourth 
is a confirmation of a recently proposed component.
Hence PSR~B0329+54 is now known to have 9 emission components 
-- the highest among all known pulsars. 
The distribution  of the pulse components around the central core 
component indicates that the emission beam consists of four nested 
cones. The asymmetry in the location of the conal components in the 
leading versus trailing parts of the profile is interpreted as being 
due to aberration and retardation in the pulsar magnetosphere. These 
measurements allow us to determine the precise location of the 4 conal 
rings of emission.  We find that the successive outer cones are emitted 
at higher altitudes in the magnetosphere.  Further, for any given cone,
the emission height at the lower frequency is found to be more than
that at the higher frequency.  
The inferred heights range from $\sim$ 160 km to $\sim$ 1150 km.
The set of ``active'' field lines, from which most of the conal 
radiation appears to originate, are found to be confined to a region
located within $\sim$ 0.5 to $\sim$ 0.6 of the polar cap radius. 
We discuss the implications of our new findings on our understanding
of the pulsar emission geometry and its impact on the emission 
mechanisms.
\end{abstract}
\keywords{pulsars: individual: PSR~B0329+54, emission beam structure, polar cap.}

\section{Introduction}
Radio wave emission from pulsars is believed to originate in the open 
field line region of the polar cap of the neutron star.  The size, 
shape and location of regions of radio emission in the average 
profiles of pulsars is thus expected to reflect the arrangement of 
emission regions in the pulsar magnetosphere.
Pulsar average profiles exhibit a great diversity in shape, and their
classification based on the number of emission components is a useful 
starting point to study the emission characteristics of pulsars. 
\cite{ra, rb, rd, rf} has carried out 
such a detailed classification, and has come up with the conclusion 
that there are two kinds of emission components -- core and conal -- 
in pulsar profiles, which result from two distinct types of emission 
mechanisms.  Further, Rankin proposes that the conal components arise 
from two nested hollow cones of emission, which along with a central 
core emission region, make up the complete pulsar emission beam.  The 
actual profile observed for a given pulsar depends on the cut that 
the observer's line-of-sight makes through this emission beam.  From 
the above work, Rankin also concludes that core radiation originates 
from very close to the neutron star surface whereas the conal radiation 
comes from regions higher up in the magnetosphere.  The outer cone is 
postulated to originate higher up in the magnetosphere than the inner 
cone, but along the same set of field lines.
However, there are other reports, (e.g. Gil 1991), which
postulate that the core and conal emissions might originate 
at very similar heights in the magnetosphere.

\cite{lm} also confirm the difference in properties between the
core and conal emissions.  However, they believe that emission within 
the beam is patchy, i.e. the distribution of component locations within 
the beam is random, rather than organized in one or more hollow cones. 
They have come to the conclusion that a single emission mechanism can 
account for both the central and outer components. Further, they propose 
that the observations are best described by a gradual change in emission 
characteristics from the axial region to the outer edge of the emission 
beam, rather than by two distinct emission processes.

In this paper, we present the results from an analysis of emission components
of PSR B0329+54.  This pulsar, one of the strongest known at 
radio wavelengths, is an excellent laboratory for a detailed study of
pulsar emission physics.  The simplest radio observations of this pulsar 
show a profile with three clearly visible components and in the classification 
scheme of \cite{rd}, it is identified as a ``triple (T)'' profile. 
The central component is thought to be due to core emission and the two 
outer components are of the conal type.  However, there has been some evidence 
for the presence of more than three distinct components for this pulsar.  For 
example, \cite{hesse} has adopted five components for a study of the relative 
intensity variations between the different components. More recently, 
\cite{ki}, from a study of fitting Gaussians to distinct emission components 
in the average profile, have found that a 
five component model does not adequately describe this pulsar's profile. 
They propose a six component model and thereby question the validity of 
the core--cone structure of the emission beam.  Hence, detection of 
the correct number of emission components and their distribution with respect 
to pulse longitude, plays a crucial part in deciding whether the pulsar emission 
beam is conal or patchy. For PSR~B0329+54, this attempt is further complicated 
by the fact that this pulsar is a mode-changer, i.e.  it switches between the 
two different modes of average profile (Lyne 1971; Hesse, Sieber and Wielebinski 
1973), during which there is an appreciable change in the relative strengths 
and the locations of the emission components. 

In this paper, we present the results from an analysis of single pulse data 
of PSR B0329+54, at two different frequencies. We have used the window--threshold 
technique, which is suitable for the detection of weak emission components in 
pulsar profiles \cite{ganga}.  In section 2, we report the detection of new 
emission components in the profile of PSR~B0329+54.  In section 3 we interpret
the results in terms of aberration and retardation effects in the pulsar 
magnetosphere and use this to estimate the emission heights and polar cap
locations of the cones. Section 4 discusses the implications of our 
findings on the structure of the emission beam of this pulsar.

\section{Observations and Data analysis}
For our study of emission components of PSR~B0329+54, we considered 
two data sets: the first obtained at 606~MHz from the Lovell telescope 
at Jodrell Bank; and the second obtained at 325~MHz from the GMRT 
(Giant Metrewave Radio Telescope) at Khodad in India.  Table~\ref{tbl-1} 
summarizes the main observing parameters for these data sets.
The details of the observing system and the calibration procedures for 
the Lovell telescope are identical to that used by \cite{gl}. 

\begin{table}
\begin{center}
\caption{Single pulse observations of PSR~B0329+54. \label{tbl-1}}
\begin{tabular}{cccccc}
\tableline \tableline
Telescope & Date & Frequency & Pulses & Resolution & Bandwidth\\
          &         &\rm (MHz)         &           &\rm (ms) & \rm (MHz) \\
\tableline
\noalign{\smallskip}
GMRT & \rm March/April\, 1999 &\rm  325 &\rm   2100 &\rm  0.516 &\rm  16 \\
Lovell & \rm 30 August\, 1996 & \rm 606 & \rm 2500 &\rm  0.249 &\rm  40\\
\tableline
\end{tabular}
\end{center}
\end{table}

For the GMRT, which consists of 30 antennas each of 45 meter diameter 
(see \cite{swarup} for details), the data were obtained 
by incoherent addition of the dual polarization signals from 6 antennas.  
The bandwidth used was 16~MHz, divided into 256 spectral channels by the
digital back-ends (see \cite{gupta} for more details about the pulsar
mode of operation of the GMRT).  The raw data were integrated to a time 
resolution of 0.516 milliseconds before 
being recorded for off-line analysis, where the data were dedispersed
and gated to obtain the single pulse data.  During off-line analysis, 
care was taken to check the data for radio frequency interference signals 
and to filter out the power line frequency signals.  The final 2100 pulses
were collected from more than one observing session, each of which had 
the same observing parameters. For all the data sets from GMRT and Lovell, 
the profiles were aligned by defining the peak of the central component 
in the average profile as the location of zero pulse longitude.

The average pulse profiles obtained from the 325~MHz and 606~MHz data 
are shown in Fig.~1a~\&~b, respectively. For the Lovell data, the pulsar 
showed a mode change in the midst of the observations.  In the beginning,
for about 600 pulses, it was in the abnormal mode after which it switched
to the normal mode.  The dotted line curve in Fig.~1b shows the average 
profile when the pulsar was in the abnormal mode. All the profiles in 
Fig.~1 show 5 distinct emission components whose peaks 
lie in the longitude range of $-14\degr$ to $10\degr$.  However,
a study of the single pulses reveals the presence of significant 
emission outside these longitude ranges. For example, Fig.~2a shows
a sequence of about 300 pulses from the abnormal mode duration, where
occasional emission in the longitude ranges $-20\degr$ to $-16\degr$ 
and $13\degr$ to $18\degr$ can easily be seen.  This extra emission
can also be detected in the average profile made from these pulses (Fig.~2b).

Since the pulses which have emission in the longitude ranges $-20\degr$
to $-16\degr$ and $13\degr$ to $18\degr$ are less frequent,
these emissions are not clearly seen in the average pulse profiles in
Fig.~1. To enhance and easily detect emission from such regions, we have
used a ``window-threshold'' technique \cite{ganga} in which we set a window in longitude 
domain and employ an intensity threshold to select the single pulses that
go to make an average profile. We consider all those pulses which have 
emissions above the threshold within the window.  As a result of this 
averaging of selected pulses, emission components within the window improve
in signal to noise ratio compared to other parts of the profile and are 
more easily detected.  Thus, for example, by setting a window over the 
longitude range $-20\degr$ to $-16\degr$, and considering all 
those pulses which are above a $4\sigma$ intensity threshold (here, 
$\sigma$ is the {\em rms} of the intensity in the off pulse region) we 
obtained an average profile shown as profile number 8 in Fig.~3a~\&~b.  
This profile clearly shows an emission component (labeled 
component VIII in the figures) centered at about $-18\degr$ for the 
606~MHz data and at about $-20\degr$ for the 325~MHz data.  

The "uniqueness in the phase location of a component peak (within error bars)
with respect to the changes in the window position and threshold levels" was
used as a criterion for deciding whether a given pulse component was detected.
As a additional criterion, we also looked at its detection at other frequency
and whether it follows the general trend of radius-to-frequency mapping
followed by the known components.

By applying the technique repeatedly at different pulse longitude windows, 
we have determined the presence of 9 unique emission components, shown 
by the 9 average profiles in Fig.~3a\&b. The combined profile (Fig.~3c~\&~d),
obtained by adding all these 9 profiles, shows the relative location of the 
9 emission components of PSR~B0329+54. 

To test if this detection of components is unique, we have tried our 
window-threshold technique for choices of much broader longitude windows, 
such as those which enclose more than one component. In all such cases, 
our technique determines the same number and location of components. 
Furthermore, our technique detects no new components when applied to 
regions of longitude that are far away from the known on-pulse region.
This argues that our detected components are genuine.

\section{Interpretation}
Since we have identified 8 conal components which are evenly distributed 
in number about the core component, our results can be interpreted in 
terms of a set of nested cones of emission.  Figure 4a shows the 
location of the components in the form of 4 conal rings of emission around 
the core component, for the 606~MHz data. In Fig.~4b we show the pulse
phase extent of each cone at 325~MHz and 606~MHz, along with the location
of the nominal centre of the cone, taken to be the half-way point between
the corresponding leading and trailing components. We estimate the phase 
location of any given component as the mean of the distribution of 
phases of the peaks of single pulses which are selected by 
our window-thresholding technique.  We estimate an error on this component 
phase location as the standard deviation of the same distribution.
The error bar on the location (zero phase) of the core component is estimated to
be about $\pm 0.6\degr$ at both the frequencies.
We also estimated the component locations by fitting Gaussians to each 
component in Fig. 3a~\&~b, which gave the results that are same within the error 
bars with the method we have adopted.
From the plots in Fig.~4, it is clear that the 
conal components are squeezed closer to the core (and to each other) in 
the trailing part of the profile, as compared to the leading part.  This 
asymmetry is clearly reflected in the shift of the cone centers away from 
the core and towards the leading part of the profile.  This effect 
increases as we go from the inner cones to the outer ones.  Further, for 
the same cone, the effect is more pronounced at the lower frequency.

The above remarkable behavior has a simple interpretation: the radiation
beams from the outer cones are progressively bent or deflected in the forward 
direction, i.e. the direction of rotation of the pulsar.  Such a bending can 
be produced by aberration and retardation effects in the pulsar magnetosphere
(Cordes 1978). Below, we consider in detail 
the emission geometry in the 
pulsar magnetosphere and include the effects of aberration and retardation.  
Using these, we show that the emission heights as well as the transverse 
location of associated dipolar magnetic field lines on the polar cap can be 
uniquely estimated for each cone, at each frequency of observation.
We also consider the effect of magnetic field sweepback which can produce
bending of the cones in the opposite direction.  We find that, in the present
context, the effect of magnetic field sweepback is negligible compared to 
the aberration and retardation effects.

\subsection{Emission geometry}
The geometry relevant to our current understanding of emission of radio
radiation from the polar cap region of pulsar magnetospheres is illustrated
in Fig.~5. 
We start with a right-handed coordinate system originating at the center of
the star with the z-axis aligned parallel to the rotation axis $\hat{\Omega}$.
In the figure, $\hat{m}$ and $\hat{n}$ represent unit vectors along the magnetic 
axis of the pulsar and along the line of sight to the observer and they are 
inclined at angles $\alpha$ and $\zeta$, respectively, to the rotation axis.
The vectors $\hat{n}$, $\hat{m}$ and $\hat{\Omega}$ all lie in the x-z
plane when the rotation phase $\phi=0$.  In this geometry, we have
\begin{equation}
\hat{n}~~=~~\hat{z}\cos\zeta+\hat{x}\sin\zeta ~~~,
\end{equation}
\begin{equation}
\hat{m}~~=~~\hat{z}\cos\alpha+\sin\alpha (\hat{x}\cos\phi+\hat{y}\sin\phi)~~~,
\end{equation}
where $\zeta = \alpha+\beta,$ and $\beta$ is the impact angle of line-of-sight
with respect to magnetic axis.
The angle $\Gamma$ made by $\hat{n}$ with respect to $\hat{m}$ at any $\phi$
is then given by
\begin{equation}
\cos\Gamma~~=~~\cos\alpha\cos\zeta+\sin\alpha\sin\zeta \cos\phi~~~.
\end{equation}

Now consider the curve C in Fig.~5,  which represents a typical dipolar 
magnetic field line. Let P represent an emission point on this field line, 
with spherical coordinates $(r, \theta)$ centered on the magnetic axis.  
Since the tangent to the field line at P needs to be parallel to $\hat{n}$, 
we can establish the following relationship between $\theta$ and $\Gamma$ :
\begin{equation}
\tan\theta~~=~~-\frac{3}{2\tan\Gamma}\pm \sqrt{2+\left(\frac{3}{2\tan\Gamma}\right )^2}
\end{equation}
This equation has two roots, and the continuous physical solution switches
the sign of second term on right hand side from positive to negative when 
$\Gamma$ changes sign from positive to negative. For the region close to 
the magnetic axis one can approximate Eq.~(4) as $\theta \approx (2/3)\Gamma.$

Let $({\bf r}^{\rm i}_{\rm em}, \theta^{\rm i}_{\rm em})$ be the coordinates of 
the emission point for the ${\rm i}^{\rm th}$ cone, (i = 1, 2, 3 and 4 are the cone 
numbers).  In the absence of any aberration and retardation, it is assumed that 
the radiation from these points is emitted tangential to the field lines. 
However, 
the aberration due to corotation causes emission beams to  bend towards the 
azimuthal direction such that radiation is received earlier than if there were 
no rotation.  

If $\zeta$ is the angle
between the rotation axis and the line-of-sight then the distance from the 
rotation axis to the line-of-sight is $r^{\rm i}_{\rm em} \sin\zeta .$ Hence
the rotation velocity at the emission point is 
$\Omega r^{\rm i}_{\rm em} \sin\zeta .$  If $v_\parallel$ is the velocity
of particles along the field line then the total velocity of particles
${\bf v}_{\rm tot} =\hat{b}\,\,v_\parallel + \hat{\phi}\,\,
\Omega r^{\rm i}_{\rm em} \sin\zeta ,$
where $\hat{b}$ and $\hat{\phi}$ are the unit vectors in the directions
of magnetic field and rotation, respectively. 
The aberration angle $\eta^{\rm i}_{\rm ab}$  is given by
\begin{equation}
\sin\eta^{\rm i}_{\rm ab}~~=~~\frac{\Omega r^{\rm i}_{\rm em}\sin\zeta}
     {v_{\rm tot}}~~~,
\end{equation}
where $v_{\rm tot}= \vert {\bf v}_{\rm tot}\vert .$ It is more appropriate 
to assume $v_{\rm tot}\approx c$ instead of $v_\parallel\approx c,$ 
therefore,  we have
\begin{equation}
\sin\eta^{\rm i}_{\rm ab}~~=~~
\frac{\Omega r^{\rm i}_{\rm em}\sin\zeta}{c}=\frac{r^{\rm i}_{\rm em}}{r_{\rm LC}}
\sin\zeta ~~~, 
\end{equation}
where $r_{\rm LC}=c/\Omega$ is the radius of the velocity of light cylinder for the pulsar. Hence the aberration is greater for emissions arising at larger altitudes.

Our aberration angle formula (Eq.~6) differs in two respects with 
respect to the one given in the literature (e.g., Cordes 1978; Phillips 1992): 
(i) literature formula uses $\alpha$ instead of $\zeta ,$ which is correct 
only when the magnetic axis and the line-of-sight are aligned $(\beta = 0),$
and (ii) we estimate '$\sin\eta^{\rm i}_{\rm ab}$' instead of '$\tan\eta^{\rm i}_{\rm ab}$' as it avoids
$v_{\rm tot}$ exceeding c when we set $v_\parallel \approx c.$ However, the
values of aberration angle estimated using the two formulae differ by a very little, 
i.e., of order $(r^{\rm i}_{\rm em}/r_{\rm LC})^2$ for 
$\beta\ll\alpha$.

If the cones are emitted at different altitudes from the surface of the neutron
star then the conal radiation emitted at lower altitudes takes more time to reach 
the observer compared to the arrival times of those components emitted at higher 
altitudes.  In terms of pulse phase of the received radiation, this is equivalent 
to a shift of components emitted at higher altitude to earlier phase with respect 
to phases of lower altitude components. If $r^{\rm i}_{\rm em}$ is the emission 
altitude of ${\rm i}^{\rm th}$ cone, then this retardation phase shift is given by
\begin{equation}
\eta^{\rm i}_{\rm ret}~~=~~\frac{r^{\rm i}_{\rm em}}{r_{\rm LC}} ~~~ .
\end{equation}
The net phase shift due to aberration and retardation is 
$\eta^{\rm i}~~=~~\eta^{\rm i}_{\rm ab}+\eta^{\rm i}_{\rm ret},$ which gives
\begin{equation}
\sin\left(\eta^{\rm i}-\frac{r^{\rm i}_{\rm em}}{r_{\rm LC}}\right)~~=~~\frac{r^{\rm i}_{\rm em}}{r_{\rm LC}}\sin\zeta ~~.
\end{equation}
For the small angle approximation ($\eta^{\rm i}_{\rm ab} \ll 1$) , the solution of 
this equation is given by
\begin{equation}
 r^{\rm i}_{\rm em}~~\approx~~\frac{r_{\rm LC}\,\eta^{\rm i}}{(1+\sin\zeta)}~~.
\end{equation}

Both aberration and retardation shift the pulse components emitted at higher 
altitudes to earlier phases in the received pulse. Therefore, we see an asymmetry 
in the location of the leading and trailing components with respect to the centre 
of the profile.  
If $\phi^{\rm i}_{\rm l}$ and $\phi^{\rm i}_{\rm t}$ are the measured locations of 
the leading and trailing components of a cone with respect to the centre of the 
profile, then
\begin{equation}
\phi^{\rm i}_{\rm l} ~~=~~ -\phi^{\rm i} ~+~ \eta^{\rm i} ~~~;~~~~~~~~~~~
\phi^{\rm i}_{\rm t} ~~=~~ \phi^{\rm i} ~+~ \eta^{\rm i} ~~~,
\end{equation}
where $\phi^{\rm i}$ is the phase of i$^{th}$ component when the aberration and retardation
are absent. From these, the values of $\eta^{\rm i}$ and $\phi^{\rm i}$ are easily obtained :
\begin{equation}
\eta^{\rm i} ~~=~~ \frac {(\phi^{\rm i}_{\rm t} ~+~ \phi^{\rm i}_{\rm l})} {2}~~~;~~~
\phi^{\rm i} ~~=~~  \frac {(\phi^{\rm i}_{\rm t} ~-~ \phi^{\rm i}_{\rm l})} {2} ~~~.
\end{equation}

The estimate of the net phase shift $\eta^{\rm i}$, in conjunction with Eq.~(8), 
immediately yields an estimate for the emission height $r^{\rm i}_{\rm em}$ of 
the $i^{\rm th}$ cone. Next, using the phase locations $\phi^{\rm i}$ of the conal 
components, we can estimate the tangent angle $\Gamma^{\rm i}$ from Eq.~(3), 
and the angular location $\theta_{\rm em}^{\rm i}$ of the emission point from Eq.~(4). 

Further, the knowledge of emission height and angular location can be used 
to map the respective locations of the ``foot'' points of field lines associated with 
the emission region of each cone on the polar cap.  For this, we start with the equation 
of dipolar field lines
\begin{equation}
r=r_{\rm e}\sin^{2}\theta~~~,
\end{equation}
where $r_{\rm e}$ is the field line constant, which is the equatorial distance of 
a field line from magnetic axis. Applying this at the surface of the neutron star gives
\begin{equation}
\frac{1}{r_{\rm e}^{\rm i}} ~~=~~ \frac{\sin^{2}\theta^{\rm i}_{s}}{r_{\rm s}} ~~=~~ 
\frac{\sin^{2}(s^{\rm i}/r_{\rm s})}{r_{\rm s}} ~~\simeq~~ 
\frac{{s^{\rm i}}^{2}}{r^{3}_{\rm s}} ~~~,
\end{equation}
where $s^{\rm i}$ is the distance from the magnetic axis to the foot of the field line
(colatitude), as measured on the surface of neutron star with radius 
$r_{\rm s}$ $(\sim 10$~km).  The approximate version in the
above equation is for the case $\theta^{\rm i}_{\rm s} \ll 1$.
At the emission point, the dipole field equation then yields
\begin{equation}
\frac{\sin^{2}(\theta^{\rm i}_{\rm em})}{r_{\rm em}^{\rm i}} ~~=~~ 
\frac{\sin^{2}(s^{\rm i}/r_{\rm s})}{r_{\rm s}} ~~~.
\end{equation}
Under the small angle approximation we obtain the following relationship :

\begin{equation}
s^{\rm i}~~\simeq~~ s_{\rm L} \, \sqrt{\frac{r_{\rm LC}}{r^{\rm i}_{\rm em}}}
\, \sin\theta_{\rm em}^{\rm i}~~.
\end{equation}
Here $s_{\rm L} \approx (r_{\rm s}^3/r_{\rm LC})^{1/2}$ is the distance from 
the magnetic axis to the foot of the last open field line that demarcates the
polar cap region. 

Using the above formulae, we have computed, for PSR~B0329+54, the emission heights 
and the location of the relevant field lines, independently for each of the four 
cones and at each of the two frequencies.  We have used $\alpha=30\degr$ and 
$\beta = 2.1\degr$, as proposed by \cite{rf} for this pulsar.  Further, we have 
taken the fiducial centre point of the profile as the location of the core component,
assuming that it is emitted at a very small height above the neutron star surface.
Our results are summarized in Table~\ref{tbl-2} for the lower frequency (325 MHz) 
and in Table~\ref{tbl-3} for the higher frequency (606 MHz).
The second and third columns give the measured locations of the leading and trailing 
components for each cone (with respect to the location of the core component), and 
the next two columns give the inferred values for net phase shifts and tangent angles
for the leading component of each cone. 
Column 6 gives the estimates of emission height (in km) for each cone.  Column 7 
gives the estimates for the transverse location of the associated field lines
on the polar cap.  This is characterized by $s^{\rm i}_{\rm L} = s^{\rm i}/s_{\rm L}$.
In the case of PSR~B0329+54, the magnetic and rotation axes are inclined through
an angle $\alpha = 30\degr,$ therefore, the polar cap is actually ellipsoidal with 
major axis aligned in the longitudinal direction of the pulsar (Cordes 1978). 
For the normalizing purpose we consider the position of the last open filed line 
which is on the azimuthal direction,
$s_{\rm L}\approx (r_{\rm s}^3/r_{\rm LC})^{1/2}\sim 171$~m.
The errors in component location are propagated appropriately to estimate 
the errors for all other dependent parameters given in Table~\ref{tbl-2} and 
Table~\ref{tbl-3}.

The salient results from our interpretation of the emission geometry are as follows :

(i) For the same cone, the lower frequency radio radiation is emitted at a 
higher altitude than the higher frequency radiation.  This result supports the 
canonical picture of radius-to-frequency mapping.

(ii) For the same frequency, successive cones are emitted at higher altitudes
in the pulsar magnetosphere.  The heights range from a nearly a hundred kilometers for
the innermost cone to a thousand kilometers for the outermost cone or from about
0.5\% to about 3\% of the light cylinder radius.

(iii) {\it All} cones, at {\it both} frequencies appear to originate on (or around)
a narrow set of field lines, slightly more than half-way out to the edge of
the polar cap region ($\rm s^{i}/s_{L} ~ \approx ~ 0.5 - 0.6$). 
\begin{table}
\begin{center}
\caption{Observed and inferred locations of conal components at 325 MHz.
\label{tbl-2}}
\begin{tabular}{crrrrrr}
\tableline\tableline
Cone &\multicolumn{1}{c}{\phil} &\multicolumn{1}{c} {\phit} &\multicolumn{1}{c}{
\etai} &\multicolumn{1}{c} {\gammai} &
\multicolumn{1}{c}{\remi}   & \multicolumn{1}{c}{\sii} \\
No.  &\multicolumn{1}{c}{(deg)} &\multicolumn{1}{c}{(deg)} &\multicolumn{1}{c}{
(deg)}      &\multicolumn{1}{c}{(deg)}  &
\multicolumn{1}{c}{(km)}     &                   \\
\tableline
1  &  $-$5.5  $\pm$  0.44  &  4.5  $\pm$ 0.29  &  $-$0.5  $\pm$
  0.26  &  3.3  $\pm$  0.11  &  180  $\pm$  100  &  0.5  $\pm$  0.16 \\

2  &  $-$9.5  $\pm$  0.21  &  7.3  $\pm$  0.31  &  $-$1.1  $\pm$
  0.19  &  4.8  $\pm$  0.09  &  430  $\pm$  070  &  0.5  $\pm$  0.04 \\
3  &  $-$14.0  $\pm$  0.34  &  10.0  $\pm$  0.42  &  $-$2.0  $\pm$
  0.27  &  6.5  $\pm$  0.13  &  770  $\pm$  110  &  0.5  $\pm$  0.04 \\
4  &  $-$20.0  $\pm$  1.43  &  14.1  $\pm$  0.72  &  $-$3.0  $\pm$
  0.80  &  9.0  $\pm$  0.40  &  1150  $\pm$  310  &  0.6   $\pm$  0.08 \\
\tableline
\end{tabular}
\end{center}
\end{table}

\begin{table}
\begin{center}
\caption{Observed and inferred locations of conal components at 606 MHz.
\label{tbl-3}}
\begin{tabular}{crrrrrr}
\tableline\tableline
Cone &\multicolumn{1}{c}{\phil} &\multicolumn{1}{c} {\phit} &\multicolumn{1}{c}{
\etai} &\multicolumn{1}{c} {\gammai} &
\multicolumn{1}{c}{\remi}   & \multicolumn{1}{c}{\sii} \\
No.  &\multicolumn{1}{c}{(deg)} &\multicolumn{1}{c}{(deg)} &\multicolumn{1}{c}{
(deg)}      &\multicolumn{1}{c}{(deg)}  &
\multicolumn{1}{c}{(km)}     &                   \\
\tableline
1  &  $-$5.3  $\pm$  0.35  &  4.5  $\pm$  0.37  &  $-$0.4  $\pm$
  0.25  &  3.3 $\pm$  0.10  &  160  $\pm$  100  &  0.6  $\pm$  0.17 \\
2  &  $-$8.8  $\pm$  0.34  &  6.9  $\pm$  0.14  &  $-$1.0  $\pm$
  0.18  &  4.6  $\pm$  0.08  &  380  $\pm$  070  &  0.5  $\pm$  0.05 \\
3  &  $-$12.6  $\pm$  0.64  &  9.5  $\pm$  0.64  &  $-$1.6  $\pm$
  0.46  &  6.0  $\pm$  0.22  &  600  $\pm$  180  &  0.5  $\pm$  0.08 \\
4  &  $-$17.9  $\pm$  0.79  &  13.6  $\pm$  0.88  &  $-$2.2  $\pm$
  0.59  &  8.4  $\pm$  0.29  &  840  $\pm$  230  &  0.6  $\pm$  0.09 \\
\tableline
\end{tabular}
\end{center}
\end{table}

\subsection{Effect of magnetic field sweepback}
We now consider the effect of magnetic field sweepback, an effect that opposes
the aberration effect.  Since pulsars lose their rotation energy by magnetic 
dipole radiation, the magnetic dipole field experiences a torque due to such 
emissions. As a result, the field lines tend to bend in the toroidal directions, 
which is opposite to the sense of rotation. To estimate the bending angle at 
the emission altitude $r^{\rm i}_{\rm em}$ we use the following result given by 
\cite{shi}
\begin{equation}
\phi^{\rm i}_{\rm mfs}=1.2\left(\frac{r^{\rm i}_{\rm em}}{r_{\rm LC}}\right)^3 \sin^2\alpha ~~~.
\end{equation}
Using the emission altitudes given in the Table~\ref{tbl-2}, we computed the bending angle 
due to field sweepback for each cone, at each frequency, and found the sweepback is too 
small compared to the aberration and retardation phase shifts. For the outer most cone, where
the sweepback is expected be highest, we find  $\phi^{\rm i}_{\rm mfs} < 0.001\degr$ 
at 325 MHz and 0.0005$\degr$ for 610 MHz. Therefore, our assumption that the phase 
shifts of the cone centers with respect to the core position are predominantly 
due to the aberration and retardation, is validated. 

\section{Discussion}
Our results show that PSR~B0329+54 has 9 unique emission components 
(of which at least 3 have been detected for the first time).
This is the largest number of components detected for any pulsar and 
indicates that the emission geometry for this pulsar is probably quite 
complicated.  We note that our component VI, detected in the longitude 
range of $-8\degr$ to $-10\degr$, matches very well with the 
sixth component proposed by \cite{ki} as a result 
of Gaussian fits to the emission components. Our results thus confirm 
the component proposed by them.

The new components are clearly seen at 606~MHz but are less prominent
at 325~MHz.  This may be part of the known trend of reducing significance
of conal components (with respect to the core component) at lower 
frequencies.  Further, the locations of components at the two 
frequencies follow the commonly seen trend that distance from the centre 
of the profile increases at lower frequencies, generally referred to as  
the ``radius-to-frequency-mapping'' effect, (e.g. Phillips 1992; Sieber 1996).  
This also supports the argument of genuineness of the 
detected components.  Our results also appear to indicate that the newly 
detected conal components for this pulsar show up more clearly during 
the abnormal mode rather than in the normal mode. 

Our interpretation of the 9 emission components as 4 nested cones of 
emission surrounding a single core component supports the picture of core
and conal emission beams (e.g. Rankin 1983a; Oster and Sieber 1977).  
\cite{rf} has proposed 2 conal rings to explain 
the existence of pulsars with 5 components in their average profiles.  
More recently, \cite{mitra} have found evidence for 
3 nested conal rings, from their analysis of the emission patterns of a 
large sample of pulsars.  PSR~B0329+54 is the first pulsar for which we found
the clear evidence for as many as 4 nested cones of emission.  It 
should be interesting to see if a similar analysis for other multicomponent 
pulsar shows the presence of 4 emission cones.

Our data also presents the direct evidence for the detection of 
aberration and retardation effects in pulsar magnetospheres.  Earlier experimental 
efforts to detect evidence for aberration and retardation time delays (as well as 
magnetic field sweepback) in some pulsars (e.g. Phillips 1992) did not yield 
any positive signature. 
Recently, Malov and Suleimanova (1998) have estimated the aberration and retardation
phase shifts for components I and IV of this pulsar using the average pulse profiles
at different frequencies. 
The detection of aberration and retardation phase shifts of as much as three degrees
that we report, 
supports the view that the conal emission originates at altitudes of the order 
of several hundreds of kilometers above the neutron star surface.  
\cite{rf} reports an emission altitude of 217 km 
at 1 GHz for the main conal ring (cone number 3 in our picture) of this pulsar.  
\cite{kg1, kg2} estimate somewhat higher 
emission altitudes, in the range 300 - 500 km.  From our results, the 
corresponding emission altitude is about 600 km at 606 MHz.

Further, the fact that a given conal ring exhibits more aberration at the lower
frequency clearly shows that the lower frequency radiation originates at a
higher altitude in the magnetosphere of this pulsar. This 
radius-to-frequency-mapping effect has been proposed as an explanation 
for the commonly observed fact that the overall widths of pulsar profiles increase 
with decreasing frequency.  Our results provide a direct corroboration of this 
model, and  the result then 
makes other proposed explanations, such as chromatic aberration (or refraction) 
in the magnetosphere (e.g. Barnard and Arnos 1986) or low frequency broadening 
due to reduced beaming of the radiating particles (Kunzl et al. 1998), 
less likely to be valid.

By combining the emission heights obtained from aberration and retardation effects 
with a dipolar 
magnetic field geometry, we have been able to localize the exact emission point of 
each cone with a good degree of precision.  Perhaps the most interesting result 
of our work is the conclusion that the successive cones of emission originate at 
increasing heights in the magnetosphere, but along relatively nearby field 
lines.  This is for the first time that such a result has been obtained for any 
pulsar.  In the context of 2 cones of emission, \cite{rf} has raised this 
as a major unanswered question: whether the cones are emitted at different 
heights along the same set of field lines or are associated with different sets 
of field lines. For PSR~B0329+54, this question is now answered for all the 4 cones
of emission. We find that all the cones at a given frequency are associated with a 
set of field lines that are located at about $0.5$ - $0.6$ of the distance to the 
edge of the polar cap.  We believe that the scatter in the values (0.5 to 0.6) for 
different cones is within the limits of our estimation errors.
Further, we find that the cones at both frequencies are also associated with the 
same set of field lines, again within the error estimates.  

We note in particular that our results show clearly that the radio emission does 
{\it not} originate at or near the last open field line region of pulsar 
magnetosphere.  In the formulation we have presented in Eqs. (1) through 
(15), there is only one free parameter that can affect this conclusion: the 
value of $\alpha$, the inclination angle between the magnetic and rotation
axes of the pulsar.  Higher values of $\alpha$ would reduce the estimated emission
heights and consequently move the associated field lines closer to the edge of
the polar cap. However, there good evidence that the value of $\alpha=30\degr$ 
for this pulsar is a robust estimate (Rankin 1990; Lyne and Manchester 1988).

If the results about the emission geometry that we have obtained for pulsar B0329+54
are found to be true in general for the population of known pulsars, then it 
should prove to be a significant improvement in our general understanding of 
the emission geometry of radio pulsars.  This could provide important constraints 
for the various theories for the emission mechanism of radio pulsars.  For 
example, any successful theory would need to produce radiation at a given 
frequency from significantly different heights along the same set of field lines 
in the magnetosphere to explain the observed behavior of conal components.

\section{Conclusion}
We have used a technique based on windowing and thresholding, to detect 
the weak emission components in pulsar profiles. By applying this to the single 
pulse data of PSR~B0329+54 at 325~MHz and 606~MHz, we have detected 
three new emission components for this pulsar, and also confirmed the
presence of a new component proposed by \cite{ki}. 
We conclude that this pulsar has 9 components, which is the highest 
among all the known pulsars. 

We interpret the 9 components as being produced by 4 nested hollow cones
of emission along with a central core component of emission.  Our results
thus support the core and conal emission picture of pulsar radiation beams.

We find clear evidence that the conal components in the trailing part of
the pulsar profile are squeezed closer to the core (and to each other)
as compared to the corresponding components in the leading part of the
profile.  We interpret this as evidence for aberration and retardation effects in the
conal beams.  The observed aberration and retardation phase shifts increase from the 
inner to outer cones, and for any given cone, the effect is more at the lower frequency.

From a detailed consideration of the emission geometry in the magnetosphere,
including the effects of aberration and retardation, we are able to constrain the 
location of emission regions independently for each cone, and at each
frequency (325 \& 606 MHz). From these computations, we are able to conclude the 
following:
(i) for the same cone, the lower frequency radiation comes from a higher
altitude than the higher frequency radiation,
(ii) for the same frequency, successive outer cones originate at higher altitudes
and (iii) all cones at both frequencies
appear to originate on (or around) a specific set of field lines which are
located at about half-way out to the edge of the polar cap
region.  The inferred emission heights for the cones range from $\sim$ 160
km to $\sim$ 1150 km.

It is, probably, for the first time that the emission regions corresponding 
to the conal components for a pulsar have been located so precisely and 
unambiguously.  In particular, our conclusion that there is only a restricted
region of the polar cap that is active in producing all the conal radiation
that is observed, should prove to be an interesting constrain for 
models or theories of pulsar emission mechanisms.  

\begin{acknowledgements}
We thank Dunc Lorimer for his help in data reduction and fruitful discussions
during the initial stages of the work. We are thankful to A. G. Lyne for 
providing the Jodrell Bank data and to J. M. Rankin, A. Peyman and D. Mitra for 
useful discussions.
\end{acknowledgements}

\clearpage

\begin{figure}
\plotone{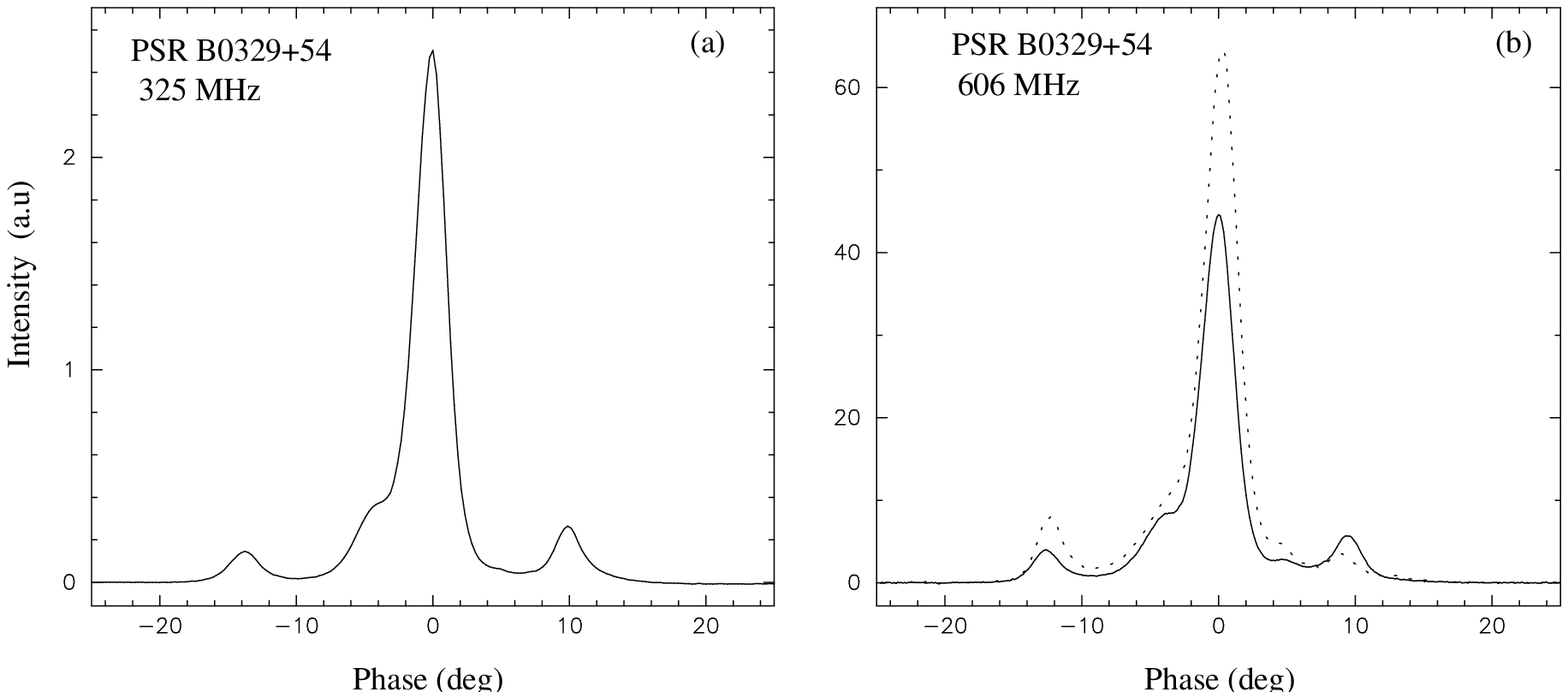}
\caption{Average pulse profiles for PSR~B0329+54 obtained from the data 
at 325~MHz and 606~MHz.  The intensity is in arbitrary units. For the 
606~MHz data, the dotted line curve shows the profile for the abnormal 
mode of the pulsar.\label{fig1}}
\end{figure}

\begin{figure}
\plotone{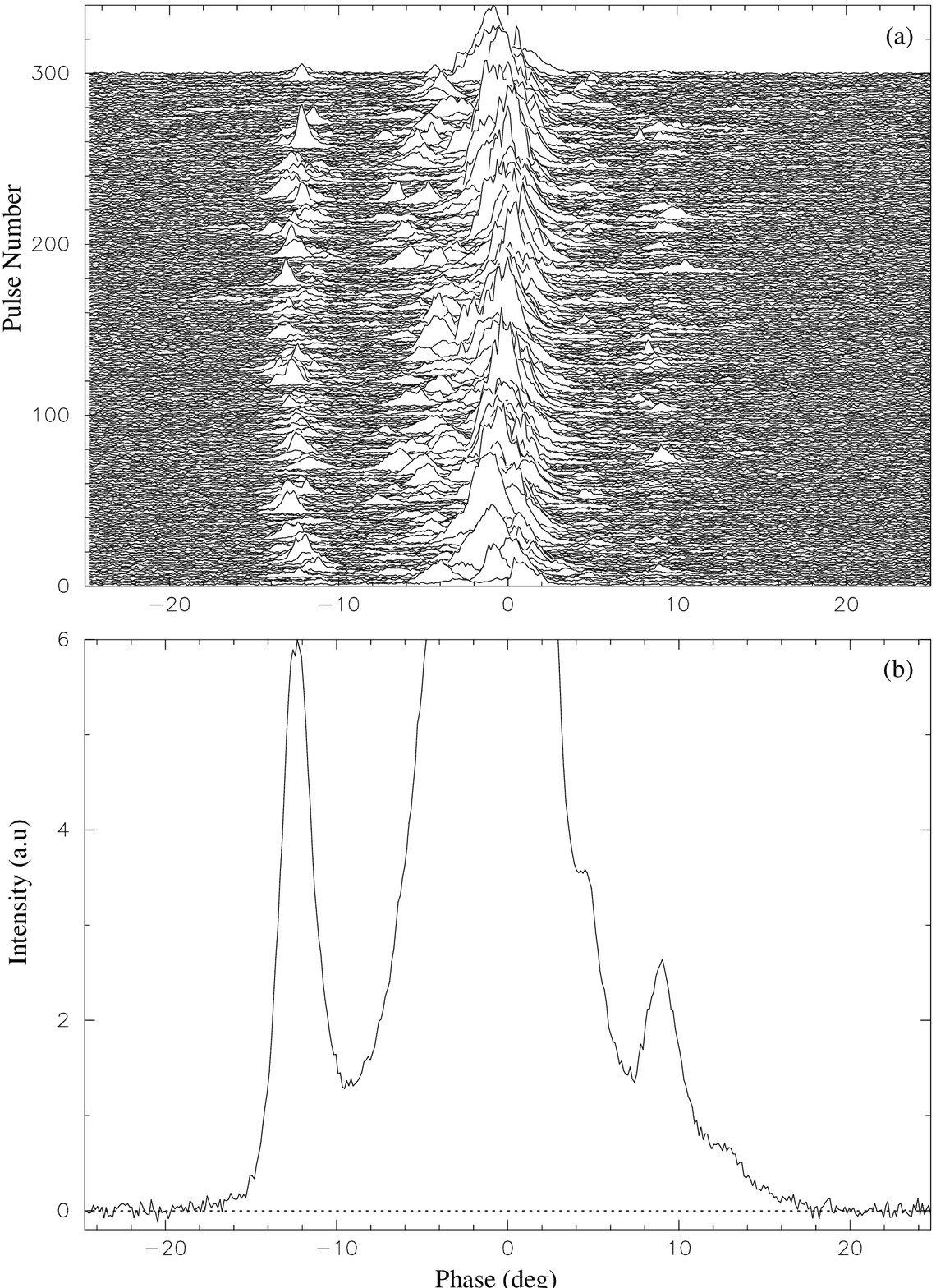}
\caption{(a) Sequence of single pulses from PSR~B0329+54 at 606~MHz, and 
(b) an average pulse profile obtained from the same, plotted with a zoomed
intensity scale to enhance the emission at the outer longitudes. The dotted
line shows the mean off pulse intensity level.\label{fig2}}
\end{figure}

\begin{figure}
\plotone{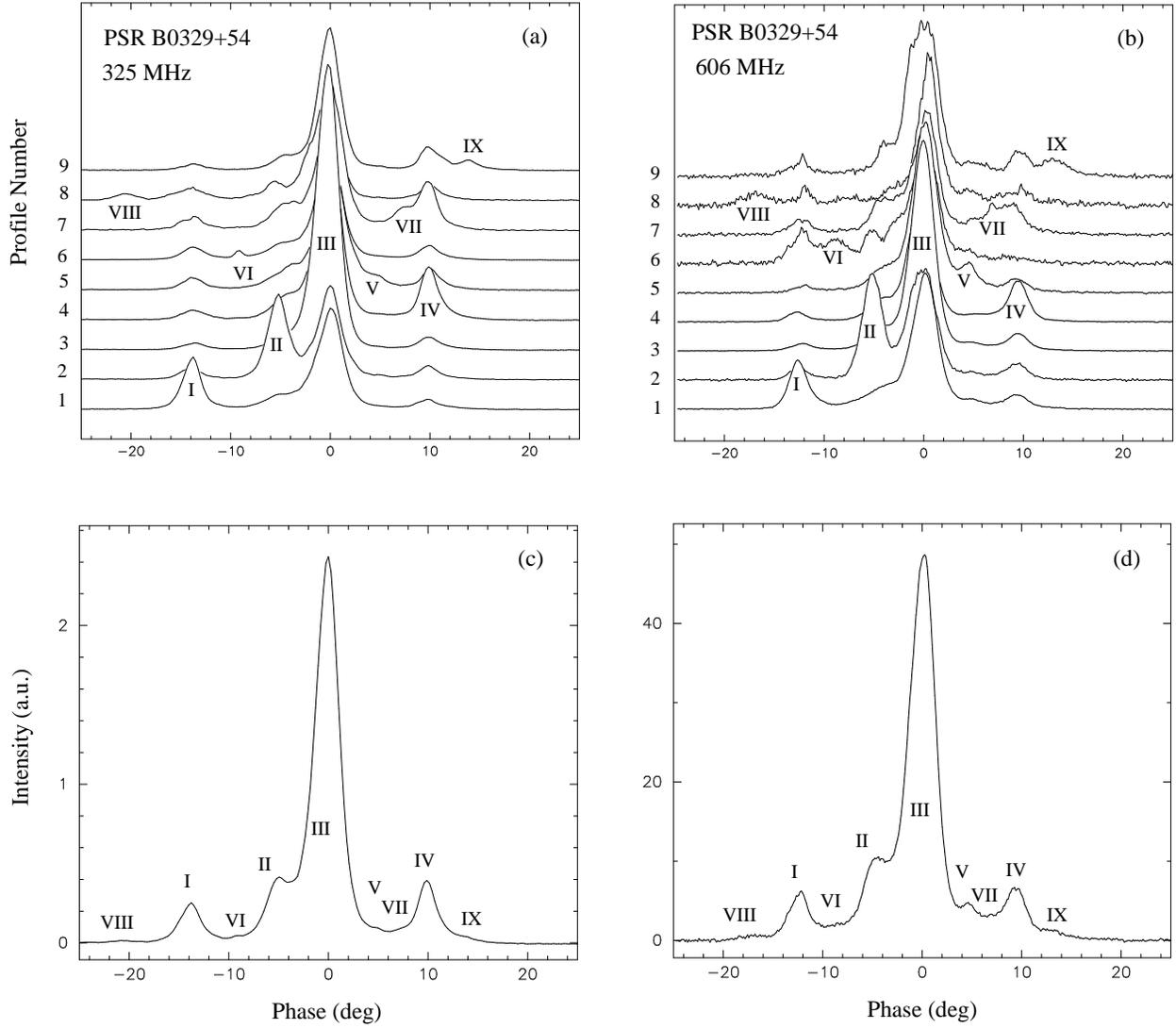}
\caption{ (a)\&(b) show the average profiles (in arbitrary units) obtained 
for each of the nine detected components by using the window-threshold 
technique at 325~MHz and 606~MHz while (c)\&(d) show the averages of these 
nine profiles.\label{fig3}}
\end{figure}

\begin{figure}
\plotone{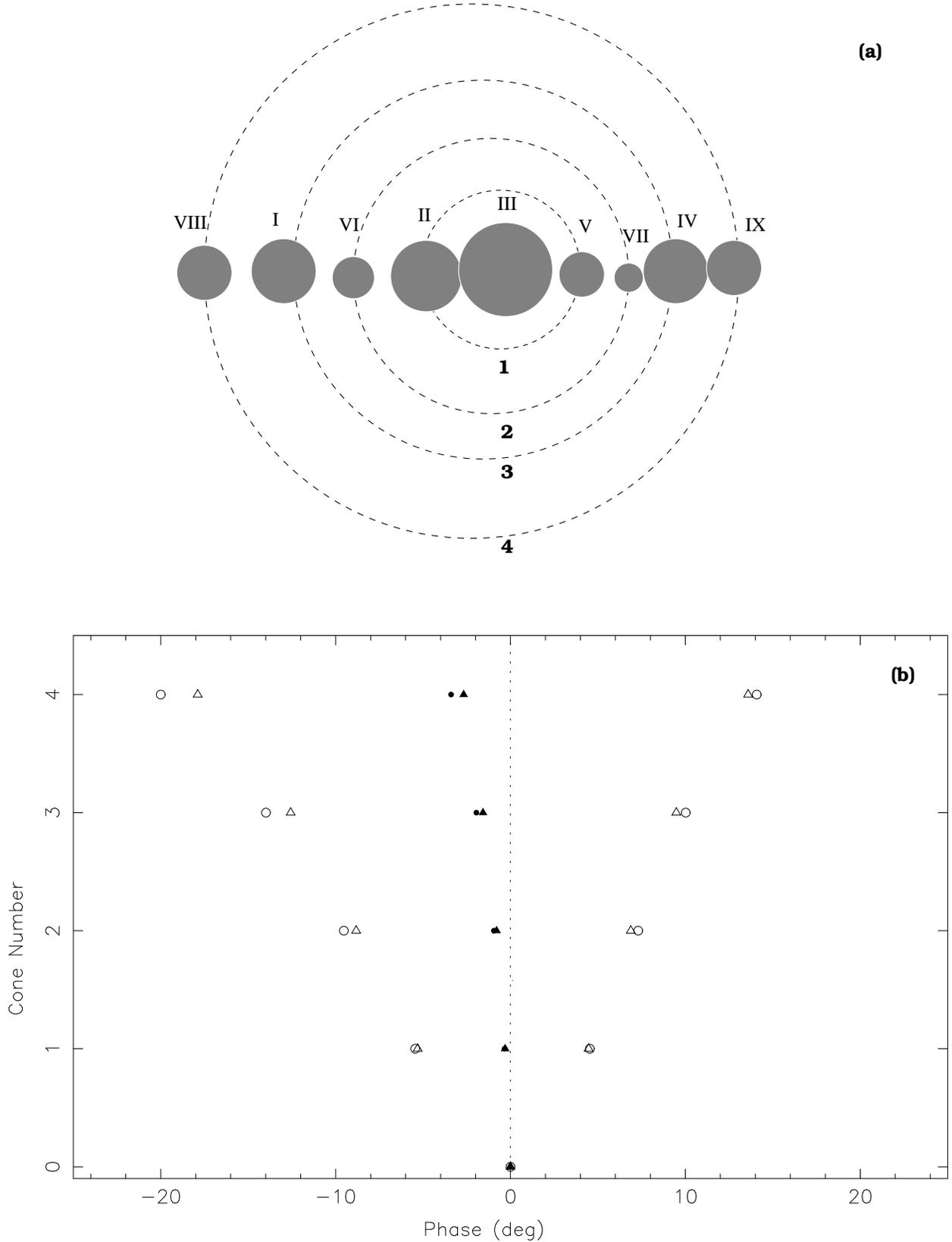}
\caption{(a) Location of the nine emission components at 606~MHz, shown in
the form of 4 conal rings around the central core component, and (b) extent
(open symbols) and centre (filled symbols) of the cones at 325~MHz (circles)
and 606~MHz (triangles).  The core component is labelled as cone number 0.
Note that the scale for (a) is same as that for (b).\label{fig4}}
\end{figure}
\vskip 5.0cm

\begin{figure}
\plotone{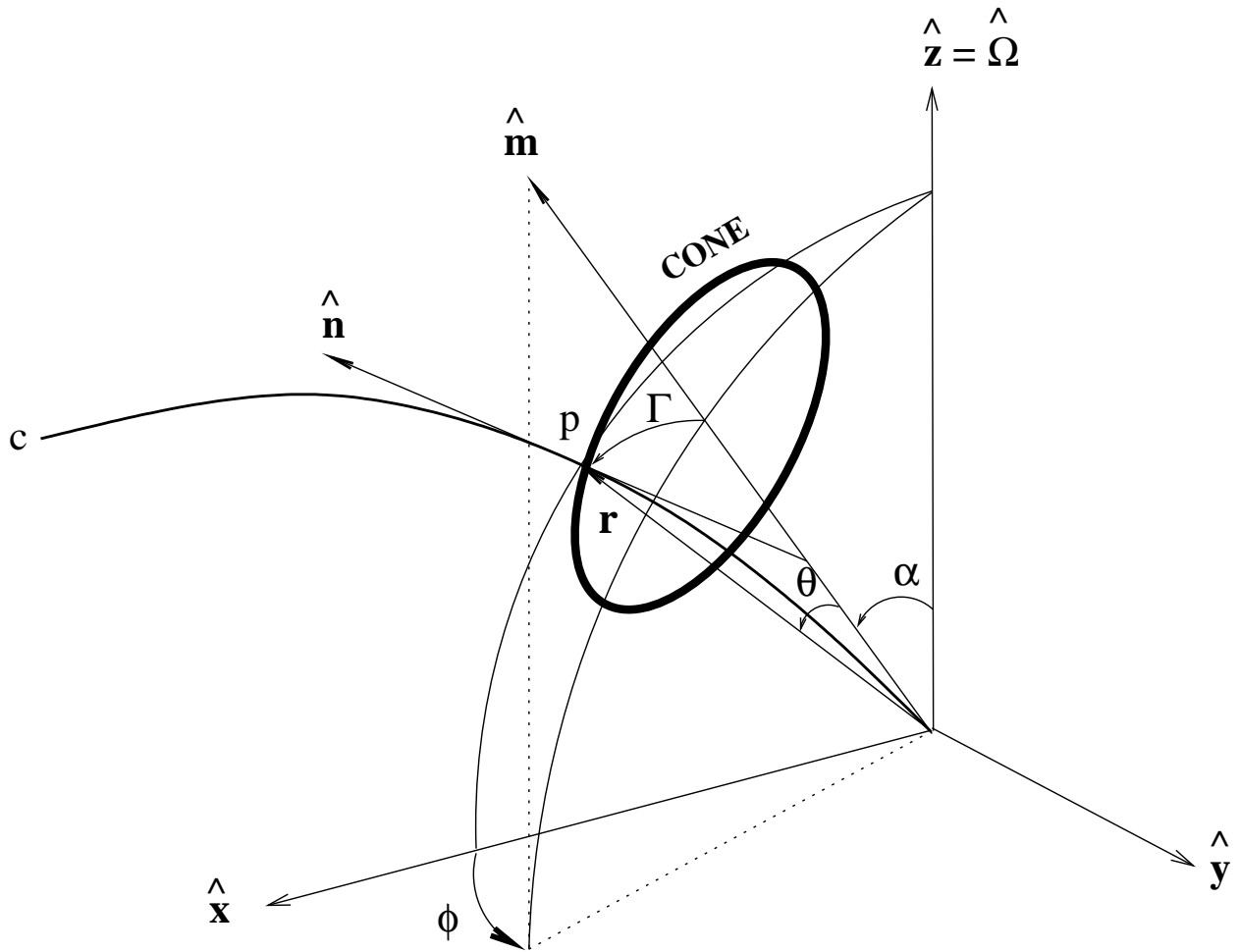}
\caption{Emission geometry, where $\hat {n}$ is tangent to the dipolar field line C
and directing towards the observer's line-of-sight, ${\bf r}$ is the position vector
of emission point p, $\hat{m}$ is the magnetic axis, and $\phi$ is the pulse phase.
\label{fig5}} 
\end{figure}
\end{document}